\documentclass[preprint,showpacs,showkeys, preprintnumbers,amsmath,amssymb,superscriptaddress, prl, linenumbers]{revtex4-1}
\usepackage{graphicx}% Include figure files
\usepackage{dcolumn}% Align table columns on decimal point
\usepackage{bm}% bold math

\begin{document}

%\preprint{APS/123-QED}

\title{Boundary Scattering in Ballistic Graphene}% Force line breaks with \\

\author{Satoru Masubuchi}
\email{msatoru@iis.u-tokyo.ac.jp}
\affiliation{Institute of Industrial Science, University of Tokyo,  4-6-1 Komaba, Meguro-ku, Tokyo 153-8505, Japan.}
\affiliation{Institute for Nano Quantum Information Electroncs, University of Tokyo, 4-6-1 Komaba, Meguro-ku Tokyo 153-8505, Japan.}

\author{Kazuyuki Iguchi}
\affiliation{Institute of Industrial Science, University of Tokyo,  4-6-1 Komaba, Meguro-ku, Tokyo 153-8505, Japan.}

\author{Takehiro Yamaguchi}
\affiliation{Institute of Industrial Science, University of Tokyo,  4-6-1 Komaba, Meguro-ku, Tokyo 153-8505, Japan.}

\author{Masahiro Onuki}
\affiliation{Institute of Industrial Science, University of Tokyo,  4-6-1 Komaba, Meguro-ku, Tokyo 153-8505, Japan.}

\author{Miho Arai}
\affiliation{Institute of Industrial Science, University of Tokyo,  4-6-1 Komaba, Meguro-ku, Tokyo 153-8505, Japan.}

\author{Kenji Watanabe}
\affiliation{National Institute for Materials Science, 1-1 Namiki, Tsukuba, 305-0044, Japan. }

\author{Takashi Taniguchi}
\affiliation{National Institute for Materials Science, 1-1 Namiki, Tsukuba, 305-0044, Japan. }

\author{Tomoki Machida}
\email{tmachida@iis.u-tokyo.ac.jp}
\affiliation{Institute of Industrial Science, University of Tokyo,  4-6-1 Komaba, Meguro-ku, Tokyo 153-8505, Japan.}
\affiliation{Institute for Nano Quantum Information Electroncs, University of Tokyo, 4-6-1 Komaba, Meguro-ku Tokyo 153-8505, Japan.}
\affiliation{PRESTO-JST, 4-1-8 Honcho, Kawaguchi, Saitama 332-0012, Japan}

\begin{abstract}
We report magnetotransport measurements in ballistic graphene/hexagonal boron nitride mesoscopic wires where the charge carrier mean free path is comparable to wire width $W$. Magnetoresistance curves show characteristic peak structures where the peak field scales with the ratio of cyclotron radius $R_\textrm{c}$ and wire width $W$ as $W/R_\textrm{c} = 0.9 \pm 0.1$, due to diffusive boundary scattering. The obtained proportionality constant between $R_\textrm{c}$ and $W$ differs from that of a classical semiconductor 2D electron system where $W/R_\textrm{c} = 0.55$. %Magnetoresistance amplitude analysis suggests nearly zero specularity at the graphene boundary.
\end{abstract}

%Use showkeys class option if keyword
                    %display DEGired
\maketitle

The presence of unique relativistic charge carriers, massless Dirac fermions, in monolayer graphene \cite{Novoselov04} has lead to the emergence of novel transport phenomena such as half-integer \cite{Novoselov05, Zhang05} and fractional quantum Hall effect \cite{Du09, Bolotin09}. When the elastic mean free path of Dirac fermions becomes comparable to the sample size, the charge carrier transport mechanism enters a quasi-ballistic regime \cite{Beenakker91}, and Dirac fermions exhibit more intriguing relativistic transport phenomena such as Klein tunneling \cite{Young09}, evanescent wave transport \cite{Tworzydo09}, and sub-Poissonian shot noise \cite{Danneau08}. In quasi-ballistic graphene devices, the probability of charge carrier scattering at sample boundaries becomes significantly larger than in the bulk region, because the probabilities of charge carriers scattering from charged impurities  \cite{Chen08A}, defects \cite{Chen09}, longitudinal acoustic phonons \cite{Chen08B, Hwang08}, substrate optical phonons \cite{Chen08B, Fratini08}, and flexural phonons  \cite{Castro10}  are significantly suppressed. To understand the fundamental transport properties of quasi-ballistic graphene, it is highly desirable to characterize the effects of boundary scattering.

In the mesoscopic wire system made from conventional semiconductor-based two-dimensional electron systems, it has been established that, if the charge carriers travel ballistically in the bulk region and are scattered diffusively at the sample boundary, the device exhibits anomalous magnetoresistance peaks owing to the magnetic commensurability effect between cyclotron radius $R_\textrm{c}$ and wire width $W$ \cite{Beenakker91, Thornton89}. Moreover, since the anomalous magnetoresistance peaks are due to diffuse boundary scattering, the probability of specular scattering, namely specularity parameter $p$, can be extracted by measuring anomalous magnetoresistance peaks \cite{Thornton89, Held99, Tokura97, Akera91}. 

Recent development of a transfer technique of graphene on hexagonal boron nitride (h-BN) \cite{Dean10, Watanabe04} have made a breakthrough in the studies of ballistic transport phenomena in graphene \cite{Mayorov11}. Since the surface of h-BN is atomically flat and h-BN has high optical phonon energy as compared to SiO$_2$, graphene on h-BN exhibits extraordinary high mobility $\mu \sim 100,000$ cm$^2$$/$Vs and a long mean free path $l_\textrm{mfp}=(h/2e) \mu \sqrt{n/\pi} \sim 1$ ${\rm \mu}\textrm{m}$ \cite{Mayorov11, Footnote01}. In conventional graphene on SiO$_2$ devices, $\mu$ is relatively low, and $l_\textrm{mfp}$ is  limited to 100 nm \cite{Chen08B}. Suspended graphene devices show high $\mu =200,000$ $\textrm{cm}^2/\textrm{Vs}$ \cite{Du08, Bolotin08}, but the observation of boundary scattering effect has not been reported \cite{Footnote02}.

In this letter, we report on magnetotransport measurements in ballistic graphene/hexagonal boron nitride mesoscopic wires, in which the carrier mean free path is comparable to the wire width. We observed anomalous magnetoresistance curves with characteristic peak structures where the peak field scales with the ratio of cyclotron radius $R_\textrm{c}$ and wire width $W$ as $W/R_\textrm{c} = 0.9 \pm 0.1$, which indicates the detection of diffusive charge carrier scattering at the graphene boundary. The obtained proportionality constant between $R_\textrm{c}$ and $W$ contrasts that of the classical semiconductor two-dimensional electron system, in which $W/R_\textrm{c} = 0.55$. In addition, from the analysis of magnetoresistance amplitude, nearly zero specularity at the graphene boundary is suggested.

The graphene/hexagonal boron nitride (GBN) mesoscopic wire system was fabricated using the following lithography steps \cite{Dean10, Masubuchi08, Masubuchi09, Masubuchi11}. First, relatively thick h-BN crystals ($\sim10$ nm) were deposited on a Si wafer using a mechanical exfoliation technique. A monolayer graphene flake was deposited  on a spin-coated polymethylmethacrylate (PMMA) layer  and transferred on a h-BN crystal using an alignment technique under an optical microscope \cite{Dean10}. Monolayer thickness of a graphene flake was verified by Raman spectroscopy. Bar-shaped geometry was defined using standard electron-beam lithography and subsequent oxygen plasma etching. The electrical contacts were defined by electron-beam lithography followed by the evaporation of Pd ($80$ nm) and lift-off. The resist residues were removed by annealing at 300 $^\circ$C in Ar/H$_2$ (97:3) gas flow for 6 h. 

In order to study the dependence of the transport properties on channel length $L$, we fabricated three devices for $L$ = 0.6, 1.4, and 2.3 $\mu$m with channel width $W$ = 1.0 $\mu$m in a single graphene flake [inset in Fig. 1(a)]. Transport measurements were carried out using the standard lock-in technique with a small alternating current of $I_\textrm{ac}=100$ nA in a variable temperature insert at $T=1.5\sim 300$ K. A heavily doped Si substrate was used as a global back gate $V_\textrm{g}$ to tune carrier density $n=C_\textrm{g} (V_\textrm{g}-V_\textrm{Dirac} )$, where $C_\textrm{g}=1.07\times 10^{-4}$ F/m$^2$ is the gate capacitance and $V_\textrm{Dirac}$ is the value of $V_\textrm{g}$ at the charge neutrality point \cite{Novoselov05}. A magnetic field was applied perpendicularly to the sample surface. 

Figure 1(a) shows conductivity $\sigma$ as a function of $V_\textrm{g}$ for the $L=2.3$ $\mu$m device measured at $T=4,16,60,120,180,240,$ and $300$ K (from top to bottom). To extract conductivity $\sigma$ from the measured resistance, the contribution of contact resistance were eliminated using transfer length method [supplementary information]. Typical V-shaped dependence of $\sigma$ on $V_\textrm{g}$ was observed for all temperatures. Minimum conductivity was located close to zero-gate bias voltages $V_\textrm{Dirac} \sim -0.3$ V, indicating small unintentional doping ($\sim 10^{10}$ cm$^{-2}$). The carrier mobility limited by long-range scattering (described below) on hole side reached $\mu_\textrm{l} = 63,000$ cm$^2$/Vs at $T = 4$ K. In the following discussions, we focus on the hole side of the $\sigma$ because the hole mobility was higher than the electron mobility. 

To extract the mean free path, we fitted the conductivity using the transport formula and assuming long- and short-range scattering as the major contributors to charge carrier scattering \cite{Hwang07, Sarma11}, $\sigma^{-1}=(ne\mu_\textrm{l} )^{-1}+\rho_\textrm{s}$, where $\mu_\textrm{l}$ is the mobility from long-range scattering, and $\rho_\textrm{s}$  is the resistivity from short-range scattering, which are both independent of charge carrier density $n$ \cite{Dean10}. In $-35$ V $<V_\textrm{g}<0$ V, the experimental data are well fitted to the calculated conductivity [dashed curve in Fig. 1(a)]. In $-50$ V $<V_\textrm{g}<-35$ V, the calculated conductivity deviates from the experimental data. This deviation can indicate the presence of an additional scattering mechanism, which is neither long- nor short-range scattering. 

From the obtained values of $\mu_\textrm{l}$ and $\rho_\textrm{s}$ for varying temperatures $T$, we evaluated the mean free path for long-range scattering $l_\textrm{long}=\hbar \sqrt{n\pi}\mu_\textrm{l} e^{-1}$ and the mean free path for short-range scattering $l_\textrm{short}=\hbar \sqrt{\pi n^{-1}} \rho_\textrm{s}^{-1}e^{-2}$ \cite{Sarma11}. The values of $l_\textrm{long}$ and $l_\textrm{short}$ as a function of $V_\textrm{g}$ for $T=4,16,60,120,180,240,$ and $300$ K are shown in Figs. 1(b) and (c) (from top to bottom). The value of $l_\textrm{long}$ increased with $|V_\textrm{g}|$ [Fig. 1(b)], whereas the value of $l_\textrm{short}$ decreased with $|V_\textrm{g}|$ [Fig. 1(c)]. As the temperature increased, both $l_\textrm{long}$ and $l_\textrm{short}$ decreased. These observations were qualitatively consistent with the previous transport measurements in graphene on SiO$_2$. Quantitatively, however, the obtained values of $l_\textrm{long}\sim1.2$ $\mu$m and $l_\textrm{short} \sim 2$ $\mu$m for $V_\textrm{g} = -45$ V were substantially larger than the typical values in graphene on SiO$_2$, which were around hundred-nanometer range \cite{Chen08B}; moreover, these values were larger than sample width $W$.

Figure 2(a) shows two-terminal resistance $R$ as a function of magnetic field $B$ at $V_\textrm{g}=-45$ V and $T=4$ K for the $L = 2.3$ $\mu$m and $W$ = 1.0 $\mu$m device. As we increased $B$ from zero, $R$ increased with $B$ and exhibited a maximum at $B_\textrm{max}=\pm 0.21$ T. As $B$ was further increased, $R$ decreased with $B$ and exhibited a minimum at $B_\textrm{min}=\pm 0.51$ T. For larger $B$, $R$ oscillated as a function of $B$, which can be attributed to Shubnikov-de Haas oscillation.

In standard theory for ballistic transport phenomena in two-dimensional electron systems in a small magnetic field \cite{Beenakker91}, if the cyclotron radius is larger than the sample width, $R_\textrm{c}>W$, and if the electrons are scattered diffusively at the sample boundary, the boundary scattering effect leads to an increase in the backscattering probability as schematically shown by solid curves in the inset of Fig. 2(a). When the magnetic field is increased, the cyclotron diameter becomes smaller than the sample width $2R_\textrm{c}<W$. Under this condition, a skipping orbit of charge carriers at the sample boundary is formed, and the backscattering probability is suppressed as schematically shown by the dashed curves in the inset of Fig. 2(a). Moreover, for this condition, the quantization of the cyclotron orbit, i.e., Landau quantization, is allowed because $2R_\textrm{c}<W$, thus Shubnikov-de Haas oscillation can be observed. The observation of anomalous peak structures and Shubnikov-de Haas oscillation in Fig. 2 is thus qualitatively consistent with the expectations from the ballistic transport with diffusive boundary scattering in the mesoscopic wire system \cite{Footnote03}. 

We studied $R$ as a function of $B$ for shorter channel lengths of $L=1.4$ and $0.6$ $\mu$m [Fig. 2(b)]. The amplitude of anomalous magnetoresistance curve, $\Delta R=R_\textrm{max}-R_\textrm{min}$ [indicated by solid and dotted arrows in Fig. 2(a), respectively], decreased with decreasing channel length $L$ [Fig. 2(c)]. In standard theory for diffusive boundary scattering, the amplitude of anomalous magnetoresistance peaks is expected to decrease with $L$, because the probability of back scattering decreases. Therefore, the observed decrease in the amplitude of magnetoresistance peaks for $L=1.4$ and $0.6$ $\mu$m devices is also consistent with the expectations from boundary scattering in the mesoscopic wire system. From the observations in Fig. 2, it can be concluded that we have observed magnetoresistance peak structures due to diffusive boundary scattering in a ballistic graphene mesoscopic wire system.

For quantitative understanding of the observed magnetoresistance curves, we discuss here the positions of magnetoresistance peaks $B_\textrm{max}$. For diffusive boundary scattering, the value of $B_\textrm{max}$ scales with the ratio of cyclotron radius $R_\textrm{c}(B, n)$ to the channel width $W$ as $W/R_\textrm{c}(B, n)=\alpha$, where $\alpha$ is the proportionality constant. For Dirac fermions in monolayer graphene, the cyclotron radius can be written as $R_\textrm{c}(B, n)=\frac{\hbar}{eB} \sqrt{\pi n}$, where $e$ is an elementary charge and $\hbar$ is Planck's constant. Therefore, the expected positions of resistance peaks in graphene can be expressed as
\begin{eqnarray}
 B_\textrm{max}=\alpha \frac{\hbar}{e} \frac{\sqrt{\pi n}}{W} 
\end{eqnarray}

To study the relationship between $R_\textrm{c}$ and $W$, we measured magnetoresistance curves at varying gate-bias voltages for the sample with $(L,W) = (2.3$ $\mu$m$, 1.0$ $\mu$m$)$. Fig. 3(a) shows two-terminal resistance $R$ as a function of magnetic field $B$ at $V_\textrm{g}=-50, -48, \cdots,-28$ V (bottom to top). In this plot, each curve was offset vertically so that peak positions were separated by 0.01 k$\Omega$. At $V_\textrm{g} = -50$ V (bottom curve), we observed the anomalous peak structures at $B_\textrm{max} = \pm 0.21$ T. As we increased gate-bias $V_\textrm{g}$, the positions of peak structures were shifted to smaller $B$. At $V_\textrm{g} = -28$ V (top curve), the peak positions were located at $B_\textrm{max} = \pm 0.14$ T. The peak positions were compared with the calculated $B_\textrm{max}$ using equation (1). As shown by the colored areas in Fig. 3(a), the peak positions were well fitted with the calculated positions  for $\alpha = 0.9 \pm 0.1$.

To investigate the dependence of $\alpha$ on the channel width, we studied another sample for larger channel width $(L,W) = (3.1$ $\mu$m$, 1.5$ $\mu$m$)$ [Fig. 3(b)]. Compared with the device with smaller $W$ [Fig. 3(a)], the values of $B_\textrm{max}$ were decreased [Fig. 3(b)]. Moreover, the peak positions were also fitted by the equation (1) for $\alpha = 0.9 \pm 0.1$, as indicated by the colored area in Fig. 3(b). These analysis were also valid for other samples for $(L,W)$ = $(1.4 $ $\mu$m$,1.0$ $\mu$m$)$  and $(2.5$ $\mu$m$,0.95$ $\mu$m$)$. These observations indicate the independence of the proportionality constant $\alpha$ on channel width $W$. 

The obtained proportionality constant $\alpha$ was larger than that of classical massive charge carriers in the conventional semiconductor two-dimensional electron system (2DES) where $\alpha = 0.55$ \cite{Thornton89}. For comparison, we plot the expected positions of peak structures for $\alpha = 0.55$ (blue curve) in Figs. 3(a) and 3(b). The curves for $\alpha =0.55$ could not explain the position of peak structures. The conventional proportionality constant between $R_\textrm{c}$ and $W$ has to be modified to explain the observed transport phenomena in graphene. 

The differing value of $W/R_\textrm{c}$ for graphene compared to that of conventional semiconductor 2DES indicate that the scattering process of massless Dirac fermions at sample boundaries can be different from that of conventional 2DEGs. As a possible explanation for the differing values of $W/R_\textrm{c}$, the interference of charge carriers at sample boundaries  can be pointed out \cite{Rakyta10}. Since the charge carriers in graphene are chiral, when the charge carrier scattering at sample boundaries changes the momentum, the phase of their wave function are shifted.  This phase shift of wave function at sample boundaries induces complex interference of charge carriers, which can alter the transport properties of ballistic graphene \cite{Rakyta10}. Moreover, the first-principle numerical calculations of the magnetoresistance in graphene nanoribbon suggested $W/R_\textrm{c} = 0.8 - 1.0$ \cite{Xu12}, which seems be consistent with our results. Albeit, for the exact understandings of the reason for the differing values of $W/R_\textrm{c}$, more elaborated works will be needed in the future. 

Finally, we study the magnetoresistance curves for varying temperatures $T$ and gate-bias voltages $V_\textrm{g}$. Fig. 4(a) shows the amplitude of magnetoresistance $\Delta R$, long-range mean free path $\l_\textrm{long}$, and short-range mean free path $l_\textrm{short}$ as a function of $T$ at $V_\textrm{g} = -35$ V. As $T$ was increased from $4$ K to $240$ K, the value of $\Delta R$ was decreased, but was retained to $T=240$ K [Fig. 4(b)]. This observation implies that the channel region of graphene mesoscopic wire was still ballistic at $T=240$ K, because of the relatively small electron-phonon scattering probability in monolayer graphene compared to that of conventional semiconductors \cite{Chen08B}, which demonstrate the characteristic electronic properties of graphene.

Fig. 4(b) shows $\Delta R$, $l_\textrm{long}$, and $l_\textrm{short}$ as a function of $V_\textrm{g}$ at $T=4$ K. The value of $\Delta R$ increased with $|V_\textrm{g}|$ from $20$ to $40$ V and saturated for $40$ to $50$ V [Fig. 4(b)]. In principle, $\Delta R$ increases with the mean free path in the bulk region, and $\Delta R$ saturates as the mean free path in the bulk region exceeds the sample width \cite{Beenakker91}. In Fig. 4(b), as we increased $|V_\textrm{g}|$, $l_\textrm{long}$ increased, and for $40$ V $<|V_\textrm{g}|<50$ V, $l_\textrm{long}$ exceeded the sample width [indicated by the dotted line in Fig. 4(b)]. On the other hand, $l_\textrm{short}$ was decreased with $|V_\textrm{g}|$ and were larger than $W$. The standard theory for ballistic transport suggests that, if specularity parameter $p$ is zero, $\Delta R$ emerges when the carrier mean free path in the bulk region becomes comparable to the sample width $W$  \cite{Thornton89}. If we assume that the charge carrier scattering in the bulk region is dominated by long-range scattering, the emergence of $\Delta R$ at $l_\textrm{long}$ $\sim 1$  $\mu$m indicates the almost zero $p$ in our device. 

In summary, we conducted magnetotransport measurements in quasi-ballistic graphene/hexagonal boron nitride mesoscopic wires, in which the carrier mean free path is comparable to the wire width. Magnetoresistance curves show characteristic peak structures where the peak field scales with the ratio of cyclotron radius $R_\textrm{c}$ and electronic wire width $W$, which indicates the detection of a diffusive scattering effect at the sample boundary. The analysis suggests that the values of $R_\textrm{c}$ and $W$ scale as $R_\textrm{c}/W=0.9 \pm 0.1$. This is in contrast to the semiconductor two-dimensional electron system where $R_\textrm{c}/W = 0.55$. The analysis also suggests nearly zero probability for specular scattering at the sample boundary. These findings are fundamental step forward in understanding the effects of boundary scattering on the transport properties of nanostructured graphene devices such as graphene nanoribbons \cite{Han07} and graphene single electron transistors \cite{Ponomarenko08, Stampfer08}. 

The authors acknowledge R. Moriya, S. Morikawa, K. Takase, and K. Muraki for technical support and helpful discussions. This study was supported by the PRESTO, Japan Science and Technology agency, the Grants-in-Aid from the MEXT, the CREST, Japan Science and Technology Agency, and the Project for Developing Innovation Systems of the MEXT.

\clearpage
\begin{figure}[t]
\begin{center}
\includegraphics{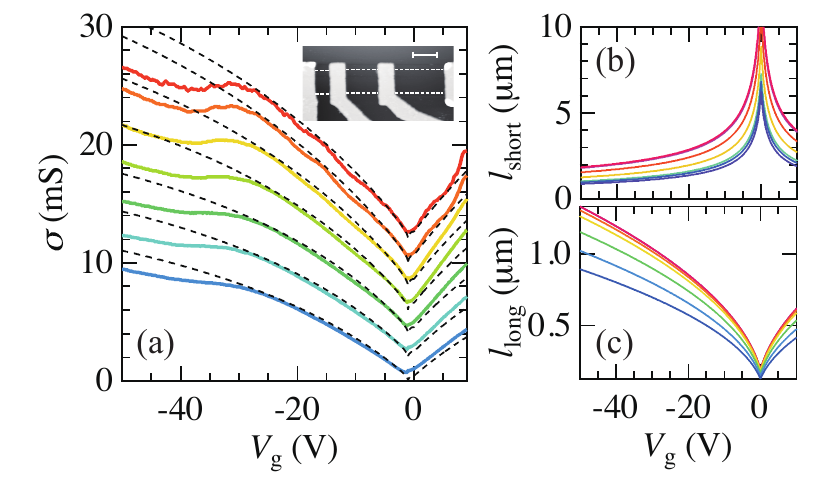} 
\caption {(Color online) (a) Conductivity $\sigma$ as a function of back-gate bias voltage $V_\textrm{g}$ measured at $T = 4, 16, 60, 120, 180, 240,$ and $300$ K (top to bottom). For clarity, each curve was offset by $2$ mS. Dashed curves are the calculated conductivity based on the Boltzmann equation discussed in the main text. The inset shows the atomic force microscopy image of the GBN sample studied in this work. The white region indicates the region of Pd metal electrodes. The white dashed lines indicate the outline of the graphene flake. (b) Short-range mean free path $l_\textrm{short}$ and (c) long-range mean free path $l_\textrm{long}$ as a function of gate-bias voltage $V_\textrm{g}$, for $T = 4, 16, 60, 120, 180, 240,$ and $300$ K (top to bottom).}
\end{center}
\end{figure}

\clearpage
\begin{figure}[t]
\begin{center}
\includegraphics{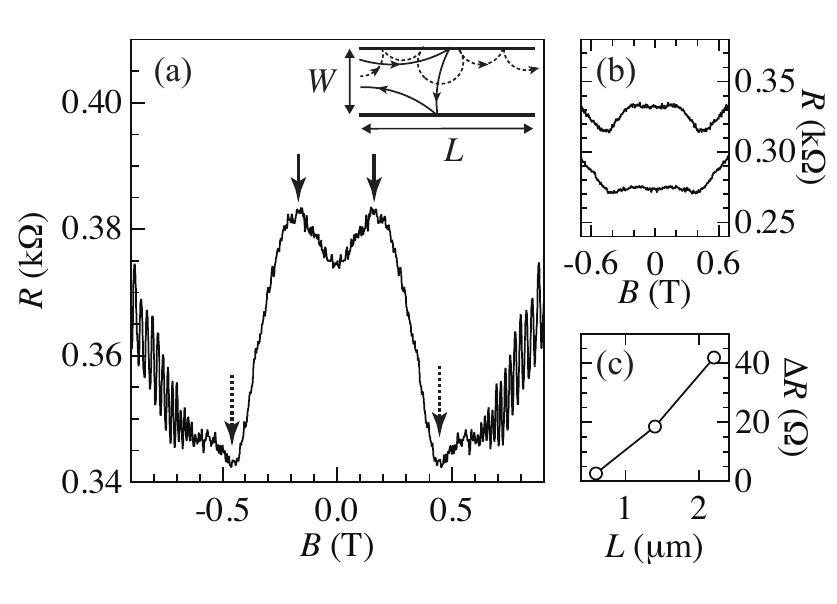} 
\caption {(a) Two-terminal resistance $R$ as a function of $B$ measured at $T = 4$ K for the device with channel length $L=2.3$ $\mu$m with applied gate-bias voltage $V_\textrm{g}=-45.0$ V. (inset) The schematic of electro-trajectories in mesoscopic wire for (solid curve) $R_\textrm{c}>W$ and (dotted curve) $R_\textrm{c}<W$. (b) $R$ as a function of $B$ measured at $T = 4$ K for (top curve) $L = 1.4$ $\mu$m and (bottom curve) $L = 0.6$ $\mu$m. (c) The amplitude of magnetoresistance signal $\Delta R$ vs. channel length $L$.}
\end{center}
\end{figure}

\clearpage
\begin{figure}[t]
\begin{center}
\includegraphics{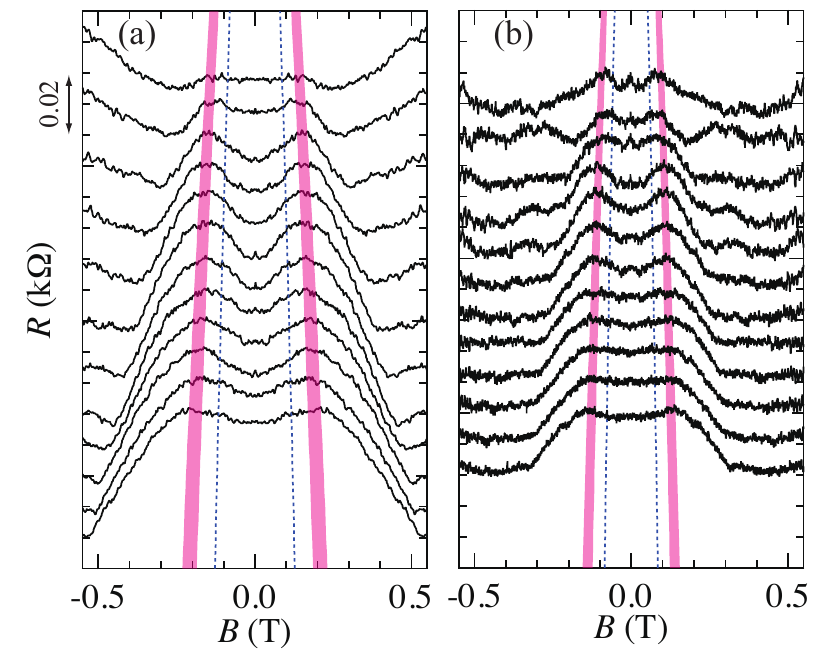} 
\caption {(Color online) (a) $R$ as a function of $B$ measured at $T = $ (a) $4$ and (b) $1.6$ K for the device with $(L, W) =  $ (a) $(2.3$ $\mu$m, 1.0 $\mu$m$)$ and (b) $(3.1$ $\mu$m, 1.5 $\mu$m$)$ with applied gate-bias voltage $V_\textrm{g}=-50, -48, \cdots, -28$ V (bottom to top). Each curve was offset vertically so that peak positions were separated by 0.01 k$\Omega$.The blue dotted curves and colored area indicate the expected peak positions for $\alpha =0.55$ and $0.9 \pm 0.1$ respectively. }
\end{center}
\end{figure}

\clearpage
\begin{figure}[t]
\begin{center}
\includegraphics{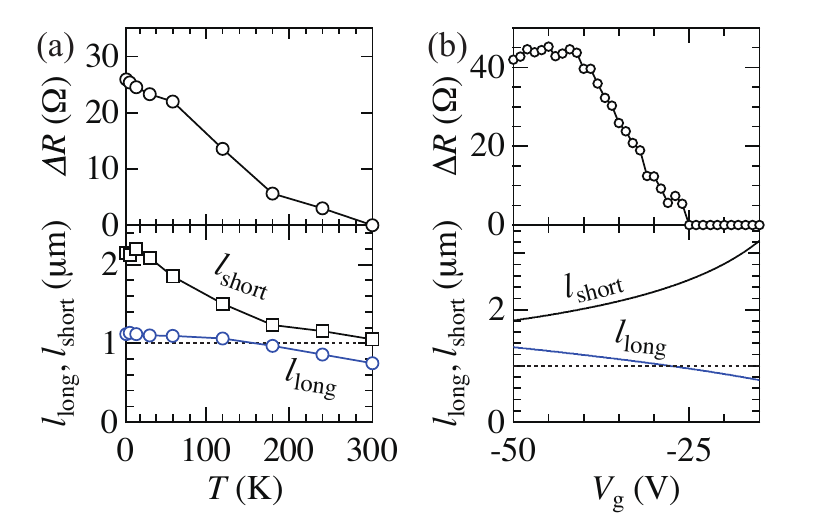} 
\caption {(a) $\Delta R$ (top panel), $l_\textrm{long}$ (blue curve in bottom panel), and $l_\textrm{short}$ (black curve in bottom panel) as a function of $T$ measured in the device with channel length $L = 2.3$ $\mu$m at $V_\textrm{g}=-35$ V. (b) $\Delta R$ (top panel), $l_\textrm{long}$ (blue curve in bottom panel), and $l_\textrm{short}$ (black curve in bottom panel) as a function of $V_\textrm{g}$ at $T=4$ K. The dotted line is the channel width of GBN mesoscopic wire.}
\end{center}
\end{figure}


\begin{thebibliography}{35}
\bibitem{Footnote01} This value was calculated for typical charge carrier concentration $n$ $\sim 10^{12}$ cm$^{-2}$
\bibitem{Footnote02} The absence of boundary scattering effect in suspended graphene might be due to limited channel length of suspended graphene.
\bibitem{Footnote03} In our measurements, the magnetic field range studied was much larger than the field range where weak localization takes place $B < 100$ mT. Therefore, we consider that the contribution of weak localization needs not be concerned in this work.
\end{thebibliography}
\end{document}